\documentclass[sn-aps]{sn-jnl}
\usepackage{graphicx}%
\usepackage{amsmath,amssymb,amsfonts}%
\usepackage{amsthm}%
\usepackage{mathrsfs}%
\usepackage[title]{appendix}%
\usepackage{xcolor}%
\usepackage{textcomp}%
\usepackage{manyfoot}%
\usepackage{booktabs}%
\usepackage{algorithm}%
\usepackage{algorithmicx}%
\usepackage{algpseudocode}%
\usepackage{listings}%
\usepackage{comment}%

\begin{document}

\title[Article Title]{SpinView: General Interactive Visual Analysis Tool for Multiscale Computational Magnetism}

\author*[1,2]{\fnm{Qichen} \sur{Xu}}\email{qichenx@kth.se}
\author*[3,4]{\fnm{Olle} \sur{Eriksson}}\email{olle.eriksson@physics.uu.se}
\author*[1,2,5]{\fnm{Anna} \sur{Delin}}\email{annadel@kth.se}

\affil[1]{
\orgname{Department of Applied Physics, School of Engineering Sciences, KTH Royal Institute of Technology}, 
\city{Stockholm}, 
\postcode{SE-10691}, 
\country{Sweden}}

\affil[2]{
\orgname{Swedish e-Science Research Center (SeRC)), KTH Royal Institute of Technology}, 
\city{Stockholm},
\postcode{SE-10044}, 
\country{Sweden}}

\affil[3]{\orgname{Department of Physics and Astronomy, Uppsala University}, 
\city{Uppsala}, 
\postcode{ Box 516, SE-75120}, 
\country{Sweden}}

\affil[4]{ \orgname{Wallenberg Initiative Materials Science for Sustainability (WISE), Uppsala University}, 
\city{Uppsala},
\postcode{SE-75120},
\country{Sweden}}

\affil[5]{ \orgname{Wallenberg Initiative Materials Science for Sustainability (WISE), KTH Royal Institute of Technology}, 
\city{Stockholm},
\postcode{SE-10044},
\country{Sweden}}

\begin{comment}
\end{comment}

\abstract{
Multiscale magnetic simulations, including micromagnetic and atomistic spin dynamics 
simulations, are widely used in the study of complex magnetic systems over a wide range of spatial and temporal scales. The advances in these simulation technologies have generated considerable amounts of data.
However, a versatile and general tool for visualization, filtering, and denoising this data is largely lacking.
To overcome these limitations, we have developed SpinView, a general interactive visual analysis tool for graphical exploration and data distillation. Combined with dynamic filters and a built-in database, it is possible to generate reproducible publication-quality images, videos, or portable interactive webpages within seconds.
Since the basic input to SpinView is a vector field, it can be directly integrated with any spin dynamics simulation tool. With minimal effort on the part of the user, SpinView delivers a simplified workflow,
speeds up analysis of complex datasets and trajectories, and
enables new types of analysis and insight.
 }

%\keywords{ToDo}

\maketitle
\section{Introduction}
Interactive postprocessing and visual exploration are essential to the analysis of ever-growing data in computational science, which serve as a conduit for facilitating knowledge mining, refining, and broadcasting to fuel interdisciplinary study\cite{stukowski2009visualization,schroeder1998visualization,momma2008vesta,ayachit2015paraview,borner2021visualizing,zheng2022skyrmion}. 
The quest to model, understand and control magnetic properties at a multitude of spatial and temporal scales -- from the quantum scale to the macroscopic scale -- is at the core of modern materials science, and necessary for the further development of technologies based on magnetic phenomena such as for example magnetic topological textures (skyrmions, hopfions), ultrafast magnetization dynamics, and magnetocalorics.
\cite{fert2017magnetic,kent2021creation,beaurepaire1996ultrafast,tegus2002transition}
To date, in the community of multiscale magnetic simulations -- with simulation packages such as UppASD\cite{UppASD_book}, Spirit\cite{bhattacharjee2012atomistic}, Vampire\cite{evans2015quantitative}, OOMMF\cite{donahue1999oommf}, and MuMax3\cite{vansteenkiste2014design} -- there is a need to simplify integration and the possibility to study the burgeoning number of emergent materials with complex magnetic properties using complementing tools.
Unfortunately, although traditional software-dependent visualization graphical user interfaces (GUI) exist, duplicated efforts such as single-use scripts and time-consuming manual analysis are still needed during the integrated use of those kinds of software (a comparison can be found in Appendix~\ref{secA1}). 
This hampers efficient knowledge sharing and creation of physical insight among researchers such as theoretical physicists and experimentalists, who typically need to minimize the time spent on scripting and other computational technicalities.

\begin{figure}[h]%
\centering
\label{Fig1}
\includegraphics[width=1\textwidth]{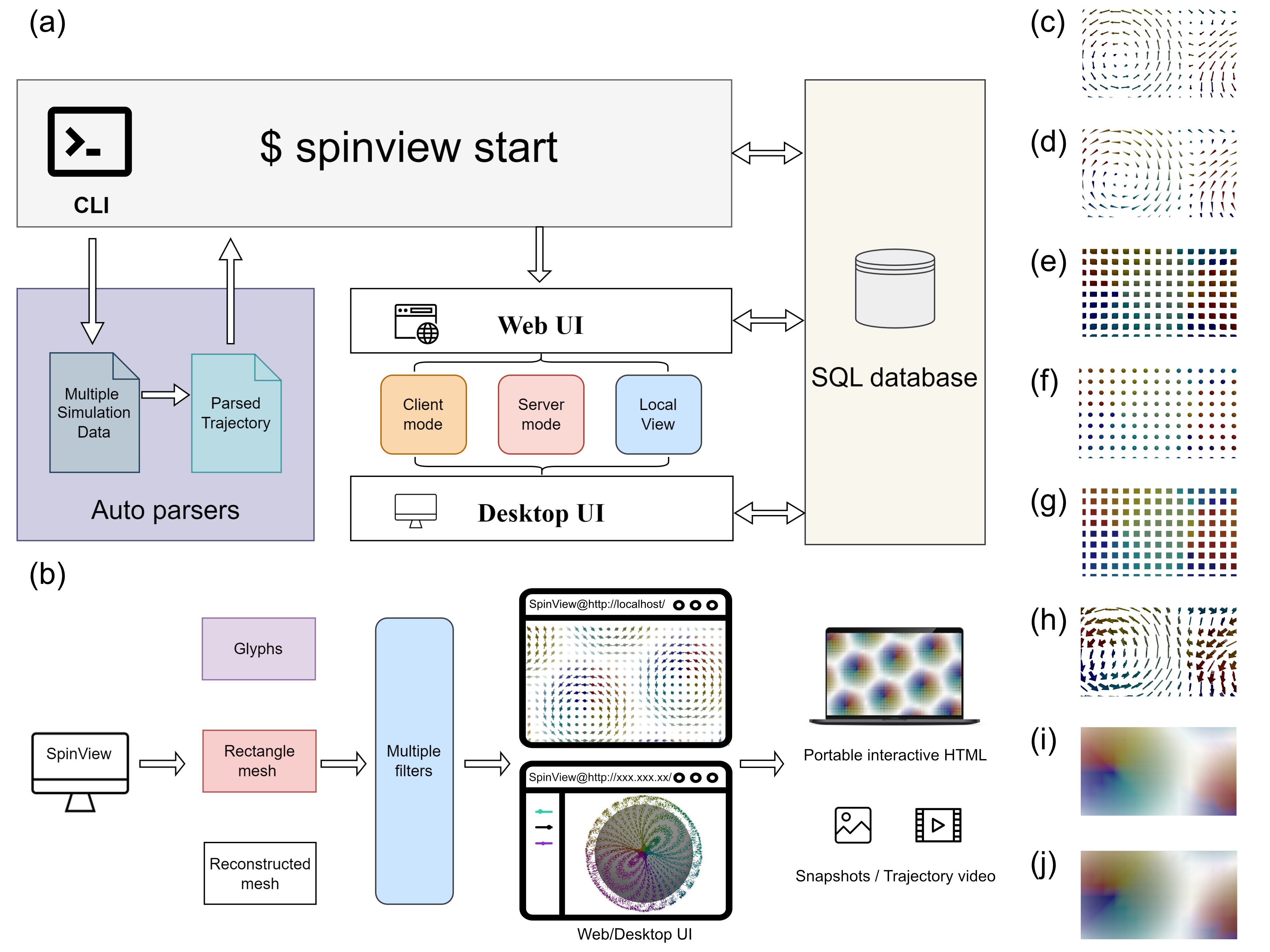}
\caption{Overview of SpinView design and workflow. (a) Four modules of SpinView: CLI, auto parser, UI, and database. (b) Basic workflow of SpinView. 
%\AnnaD{How about this:}
(c)-(h) Magnetic texture visualized by arrows, cones, boxes, spheres, planes, and a user-defined glyph. (i) The same magnetic texture as in (c)-(h), visualized using filled rectangles. The spin density at each point is computed using an interpolation scheme, which creates a smooth appearance. (j) Same as (i), but with filled triangles instead of rectangles, i.e., a triangular mesh is used. 
}\label{fig1}
\end{figure}

To handle this challenge, we report here an easy-to-use tool for interactive visual investigation and interpretation of magnetic simulation data. In this quest, we use the extendable Python programming language and leverage well-developed visualization packages from Pyvista\cite{sullivan2019pyvista} (based on the Visualization Toolkit, VTK) combined with the GUI platform Trame\cite{Trame} and CLI (command-line interface) platform Typer\cite{Typer} to develop our general visualization software SpinView. In addition to offering a variety of glyph support (including user-defined glyphs), with powerful data postprocessing ability from Numpy\cite{harris2020array}, Scipy\cite{virtanen2020scipy}, Pandas\cite{reback2020pandas}, and Pyvista, we offer easy-to-use mesh rendering and dynamic filters, e.g., advanced FFT (fast Fourier transform) denoising\cite{PhysRevB.101.184405}, projection, rescale warping, and clipping. To promote processed data sharing (with prior knowledge from professional computational scientists), we embedded a built-in SQL database that allows storing all filter parameters and provides a shareable interactive HTML file, which can directly show post-processed data in any modern web browser. Overall, SpinView is a tool that can help researchers from different research domains to obtain both nontrivial and deepened
physical insight from complex magnetic simulation results. It is also a bridge that enables seamless switching between micromagnetic and atomistic spin dynamics simulations. In this way, SpinView fills an existing gap between complex magnetic simulation data and shareable refined knowledge.

\section{Results}

\begin{figure}[h]%
\centering
\label{Fig2}
\includegraphics[width=0.8\textwidth]{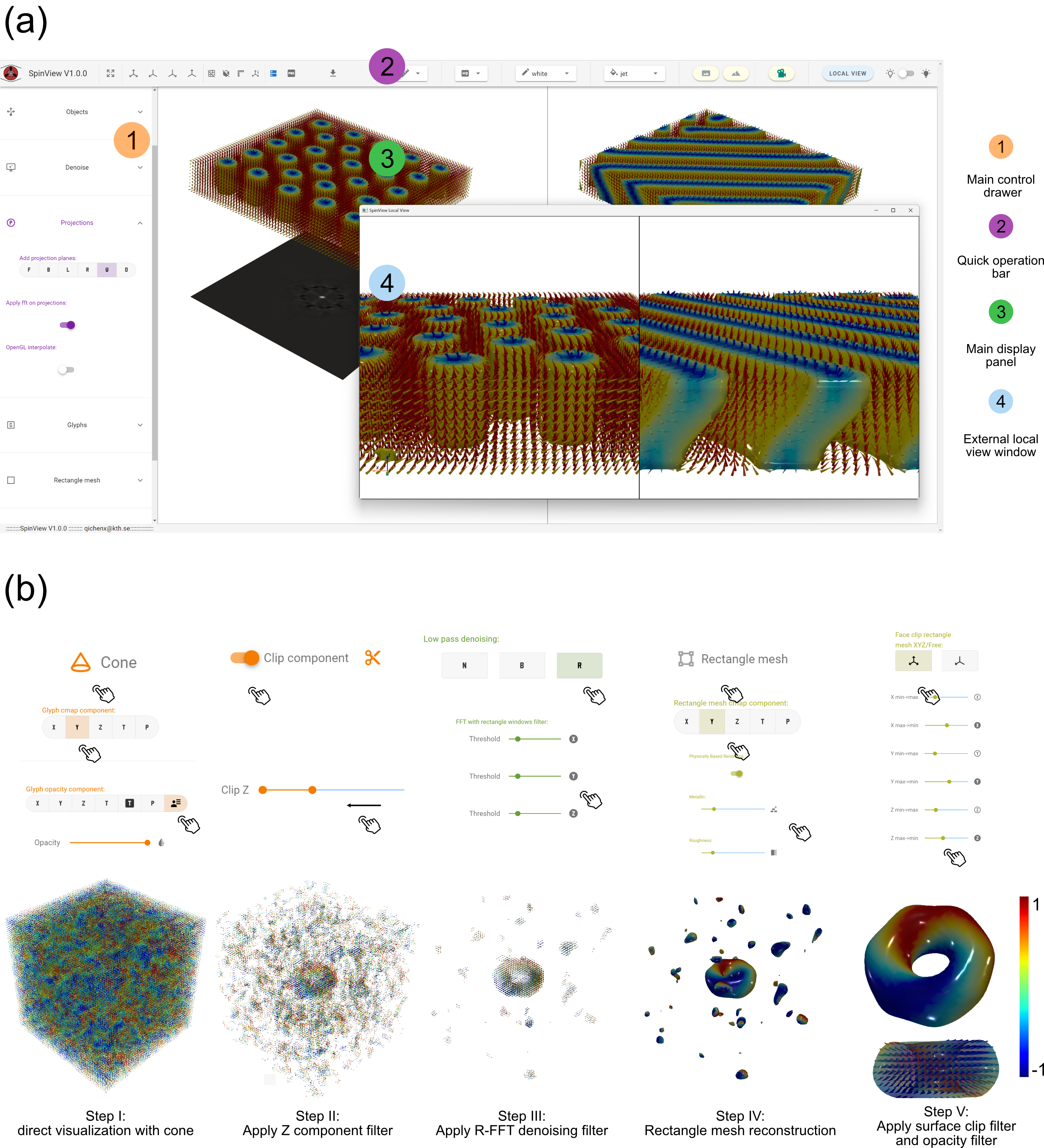}
\caption{SpinView UI and denoise workflow with interactive operation (a) General UI of SpinView. Panel 1 is the control drawer, including interactive modules, e.g., sliders, click-boxes, and drop-down menus. Panel 2 is the quick operation bar that contains often used
features, including options such as changing mode, resetting the camera, and rendering snapshots and movies. Panel 3 is the main display panel, which supports a maximum of 4 subwindows 
(See Appendix~\ref{secA2} (d)) for comparison, and two subwindows are used here. The projection plane shown in the left subwindows is the projection from the top applied with Fast Fourier Transform. Panel 4 is the external local view window, which brings smoother rendering and can pass the camera information back to the main panel when closed. (b) The data mining workflow that uses multiple filters includes rectangular FFT low pass filter.}\label{fig2}
\end{figure}

\subsection{Architecture, basic workflow and features}

Fig. 1 (a) shows four main modules in the SpinView architecture: (i) A CLI works as a global controller and main UI launcher. More information can be found in Appendix~\ref{secA2} (a)-(c). (ii) An auto parser module that automatically parses all supported kinds of input data, i.e., .out(UppASD), ASCII-based .ovf (OOMMF, Mumax, and Spirit), .csv(UppASD) and .data(Vampire) files, from different computational magnetic simulation codes into unified vector field data (in NumPy array), and organized them into trajectories in such a way as not to lose numerical efficiency. Examples can be found in Appendix \ref{secA3} (a)-(c). (iii) Main UI that includes both Web-based and desktop UI with client, server, and local view mode. (iv) A SQL database module (SQLite 3) that is used to store all users' profiles for reuse and manage data between UI and backend visualization engine. Example of this is shown in Appendix~\ref{secA4} (a).

Fig. 1 (a) and (b) show a basic visualization and postprocessing pipeline. The SpinView CLI is executed in one simulation folder. Auto parser module automatically parses all output file into vector field data and send it to the UI while the UI get all pre-set parameters from the SQL database. Multiple filters are applied under the user's operation, and the final result can be exported as portable interactive HTML files, rendered movies, or static snapshots for sharing, while a changed user profile can be stored in the SQL database. 

As shown in Fig. 2 (a), SpinView has four main panels, i.e., control drawer, quick operation bar, main display panel, and external local view windows. The keyboard assistant is globally activated for moving sliders and choosing in drop-down menus. The main display panel supports a maximum of up to four subwindows. Linked camera control is applied in all subwindows to offer a unified camera angle except in the local view mode, where 
independent camera control is optional. With support from the open source community, SpinView is highly scalable and expendable, for example with the Pyarrow backend, it can handle huge-volume trajectory, e.g., read dozens of gigabytes of CSV-based UppASD trajectory in around minutes, and it can easily add new parsers in auto parser module to support new simulation code without change other modules. Similar tools can be compared in Appendix \ref{secA1}. 

\subsection{Visual representation}

\subsubsection{Basic representation} 
As shown in Fig. 1 (c) to (j),  in SpinView, the vector field data includes snapshots and trajectories from magnetic simulations that have two kinds of representations. One is "glyph representation", e.g., arrows, cones, boxes, spheres, or planes are used to represent the data. SpinView also supports the use of dots 
or arbitrary 3D mesh data that is defined by the user. 
The other is "3D mesh", i.e., rectangle and triangular mesh representations. The rectangle mesh can efficiently describe standard rectangle mesh data or atomistic spin data with an underlying cubic crystal structure, while the triangular mesh, which is reconstructed by the Delaunay triangulation algorithm, is more general and can be used for any kinds of data but needs more processing time. Meanwhile, variable opacities are supported globally, e.g., when rendering glyphs and meshes.

\subsubsection{Simulation data cleaning and mining with built-in filter}
SpinView supports multiple filters that allow data cleaning and mining within the visualization workflow. 
We create a hopfion is a system with thermal fluctuations, in order to demonstrate this powerful feature, as shown in Fig. 2 (b) step I. Almost no valuable information can be obtained with direct visualization using the glyph cone. Then, in Step II, we applied a Z-component clip filter that filtered out all cones with a Z-component larger than -0.3. With this filter, we could find the hopfion, albeit surrounded by noise -- it is vaguely visible inside the system. To get a clear contour of the hopfion, we then applied a rectangular FFT low pass denoising filter and tuned the size of the filter window to get a minimal distortion effect and get a high-quality hopfion contour. After the denoising filter, we constructed an isosurface with the Z-component equal to 0.3 and colored it with the Y-component under Jet colormaps in step IV. Finally, we used clip filters in Step V to eliminate the still-existing noise bubbles.  Opacity was also tuned to make the generated isosurface transparent and allow the visualizable overlap between the isosurface and the glyph cone representation. With those operations, we managed to successfully reveal the topological structure despite thermal noise -- a common need for researchers interested in magnetic textures with non-trivial topology at finite temperatures. More examples can be found in Appendix ~\ref{secA3} (d) and  Appendix ~\ref{secA4} (b).

\section{Discussion}
In conclusion, SpinView is a general interactive visualization tool for computational scientists to clean, mine, and analyze multiscale  computational magnetism data. (We note that, in principle, it can of course be used to visualize any 3D vector field data.)
It is developed with researchers from several different disciplines in mind -- researchers who with minimal programming or scripting effort want to visualize data from multiscale computational magnetic 
simulations in their studies. SpinView is an open-the-box tool with rich features that can simplify magnetic data postprocessing and exploration. It can generate publication-quality figures, shareable interactive HTML files, and high-quality trajectory video in seconds with a stored profile in the built-in database. We hope and expect that SpinView will significantly increase the use of advanced visualization and filtering tools in computational magnetism and thereby enable novel findings and insight. 
\section{Methods}
\subsection{Implementation concept}
SpinView is written in Python and has three tiers in its architecture, i.e., presentation, data processing, and data tier.
The presentation tier, which includes CLI, Web UI, desktop UI, and local view, is based on the Typer, Trame, and Pyvista frameworks. All of them are cross-platform, easy to extend, and make SpinView user-friendly. The core components in the data processing tier include design filters constructed by Numpy, SciPy, Pandas, and Pyvista. A python-built-in SQLite3 database is used to store user's profiles and interact with UI through CLI.

\subsection{Multiscale computational magnetic simulation}

Micromagnetic demos shown in this work are simulated by Mumax3 and OOMMF, whereas atomistic spin dynamics data are generated by UppASD, Spirit, and Vampire.

\section{Data availability}
All data needed for reproducing the results can be found in the GitHub repository \url{https://github.com/MXJK851/SpinView/}, and the home page of the project is \url{https://mxjk851.github.io/SpinView/}. There you also find the online manual and interactive feature description. 

\section{Code availability}
All code of SpinView is available at \url{https://github.com/MXJK851/SpinView/} under the GPL-3.0 license. Interactive feature documentation can be found at \url{https://mxjk851.github.io/SpinView/}. SpinView passed the internal beta test on Windows, MacOS, and Linux and the volunteer's backgrounds include computational physics, experimental physics, and computer science from and out of the magnetism community. 

\bmhead{Acknowledgments}
This work was financially supported by the Knut and Alice Wallenberg Foundation through grant numbers 2018.0060, 2021.0246, and 2022.0108.
Q.X. acknowledges the China Scholarship Council (201906920083). O.E. and A.D. acknowledge support from the Wallenberg Initiative Materials Science for Sustainability (WISE) funded by the Knut and Alice Wallenberg Foundation (KAW). 
AD also acknowledges financial support from the Swedish Research Council (Vetenskapsrådet, VR), Grant No. 2016-05980 and Grant No. 2019-05304. O.E. also acknowledges support by the Swedish Research Council (VR), the European Research Council (854843-FASTCORR), eSSENCE and STandUP. The computations/data handling were partly enabled by resources provided by the Swedish National Infrastructure for Computing (SNIC) at the National Supercomputing Centre (NSC, Tetralith cluster) partially funded by the Swedish Research Council through grant agreement no.\,2018-05973 and by the National Academic Infrastructure for Supercomputing in Sweden (NAISS) at the National Supercomputing Centre (NSC, Tetralith cluster) partially funded by the Swedish Research Council through grant agreement no.\,2022-06725.

The authors acknowledge Filipp N. Rybakov, Alexander Edström, Vladislav Borisov, Jonathan Chico, and Mathias Augustin for valuable discussion during the code development. The authors would also like to extend gratitude to Wanjian Yin, Zhenzhu Li, Zhuanglin Shen, Qinda Guo, Zhenyang Li, Guoqiang Feng, Zhiwei Lu, Yaoxuan Zhu, I. P. Miranda, Manuel Pereiro, and Liuzhen Yang for the internal beta test and discussions.

\begin{appendices}

\section{}\label{secA1}
The comparison between the SpinView and several popular software-dependent GUIs available for magnetic simulation data visualization is shown in Table ~\ref{table1}.

\begin{table}[h]
\caption{Table1: comparison with similar tools}
\label{table1}
\begin{tabular}{@{}llllll@{}}

Functions & ASD-GUI[1]   & Spirit UI[2] & Mumax-view[3] & Vampire[4] & SpinView\\
\midrule
Web UI  & \checkmark& \checkmark &\checkmark  &  & \checkmark \\
Desktop UI  & \checkmark& \checkmark &  &  & \checkmark \\
CLI    &  &     &\checkmark& \checkmark  & \checkmark \\
Database for user's profiles   & &  &  &  & \checkmark \\
Multi-file-formats      &out &ovf  & ovf & data  & out/ovf/data \\
Alloy     & &  &  &  & \checkmark  \\
Interactive frame selection  & &  &\checkmark   &  & \checkmark  \\
Auto trajectory organize & &  & \checkmark  &  &  \checkmark \\
Snapshot generation   &  \checkmark & \checkmark &  &  & \checkmark \\
Snapshot customization  & &  &  &  & \checkmark  \\
Movie rendering & &  &  &  & \checkmark \\
Movie rendering customization& &  &  &  & \checkmark \\
Portable interactive HTML & &  &  &  &  \checkmark \\
Multiple subwindows  & &  &  &  &   \checkmark\\
Dark mode & &  &  &  &  \checkmark\\
Ruler support & &  &  &  &  \checkmark \\
Easy camera setting    & \checkmark &\checkmark  &  &  &   \checkmark\\
Multi Colormap   &4 &  6 & 1 &  & 165 \\
Multi background   &1 & 3 & 1 &  &  151\\
Opacity support & &  &  &  & \checkmark \\
Brightness support & & \checkmark & \checkmark &  &  \\
Triangular mesh reconstruction & &  &  &  & \checkmark \\
Rectangle mesh data rendering & &  &  &  & \checkmark \\
physical-based rendering support & \checkmark&  &  &  & \checkmark \\
Point support  & & \checkmark &  &  &  \checkmark\\
Baisc Glyph support  &4 &3  & 2 &  & 5 \\
Glyph property tuning  & &  &\checkmark  &  &  \checkmark\\
Arbitrary user-defined glyph    & &  &  &  & \checkmark \\
Glyph component filters & & Z &  &  & X/Y/Z/P/T\\
Surface clip Filter &X/Y/Z&  & X/Y/Z &  & X/Y/Z/A[7]\\
User-defined multiple isosurface  & &  &  &  & \checkmark \\
Isosurface components filter & &\checkmark  &  &  & \checkmark \\
Mesh data components filter & &  &  &  &  \checkmark\\
Sphere Wrapping   & &\checkmark  &  &  &  \checkmark\\
Rescaling depend on components & &  &  &  &  \checkmark\\
Rescaling ratio   & &  &  &  & \checkmark\\
Denoising support & &  &  &  &  \checkmark\\
Projection support & & \checkmark &  &  & \checkmark \\
Overlap support & &\checkmark  &  &  & \checkmark \\
Transparent isosurfaces & &  &  &  & \checkmark \\
Platform-independent   &\checkmark & \checkmark & \checkmark &\checkmark  & \checkmark \\
Open-source   &\checkmark & \checkmark & \checkmark &\checkmark  & \checkmark \\
\botrule
\end{tabular}
\footnotetext[1]{Based on the current version V1.3.}
\footnotetext[2]{Based on Spirit web UI \hyperlink{Spirit-web-ui}{https://spirit-code.github.io/web.html}.}
\footnotetext[3]{Based on current version of mumax-view web UI \hyperlink{mumax-web-ui}{https://mumax.ugent.be/mumax-view/}}
\footnotetext[4]{Based on the POV-Ray, UI for visualization is not founded}
\footnotetext[5]{Surface clip filter through arbitrary-axis (defined by the user) }

\end{table}

\section{}\label{secA2}
The example of the command line interface and user interface with four subwindows are shown in Fig. B1.

\begin{figure}[h]
\centering
\includegraphics[width=1\textwidth]{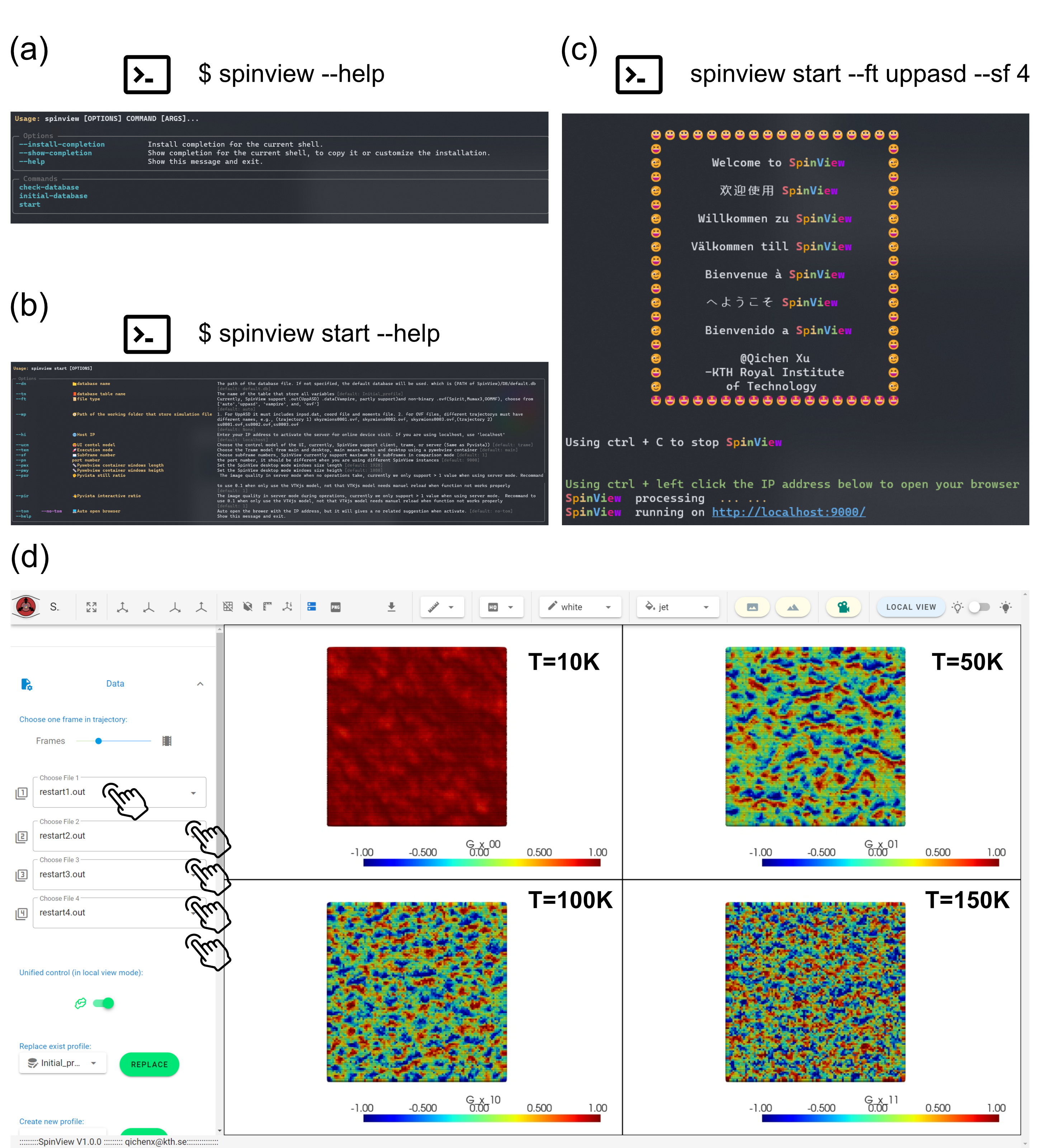}
\caption{SpinView's command line interface and user interface with four subwindows. (a) The help function interface of SpinView. (b) The help interface of 'start' functions. (c) Execute SpinView in auto mode and the welcome screen.  (d) The user interface with four subwindows shows four trajectories of the same system at different temperatures.}\label{secA2_figure}
\end{figure}

\section{}\label{secA3}
The example of using SpinView with multiple datatypes and visualization types is shown in Fig. C2.
\begin{figure}[h]
\centering
\includegraphics[width=1\textwidth]{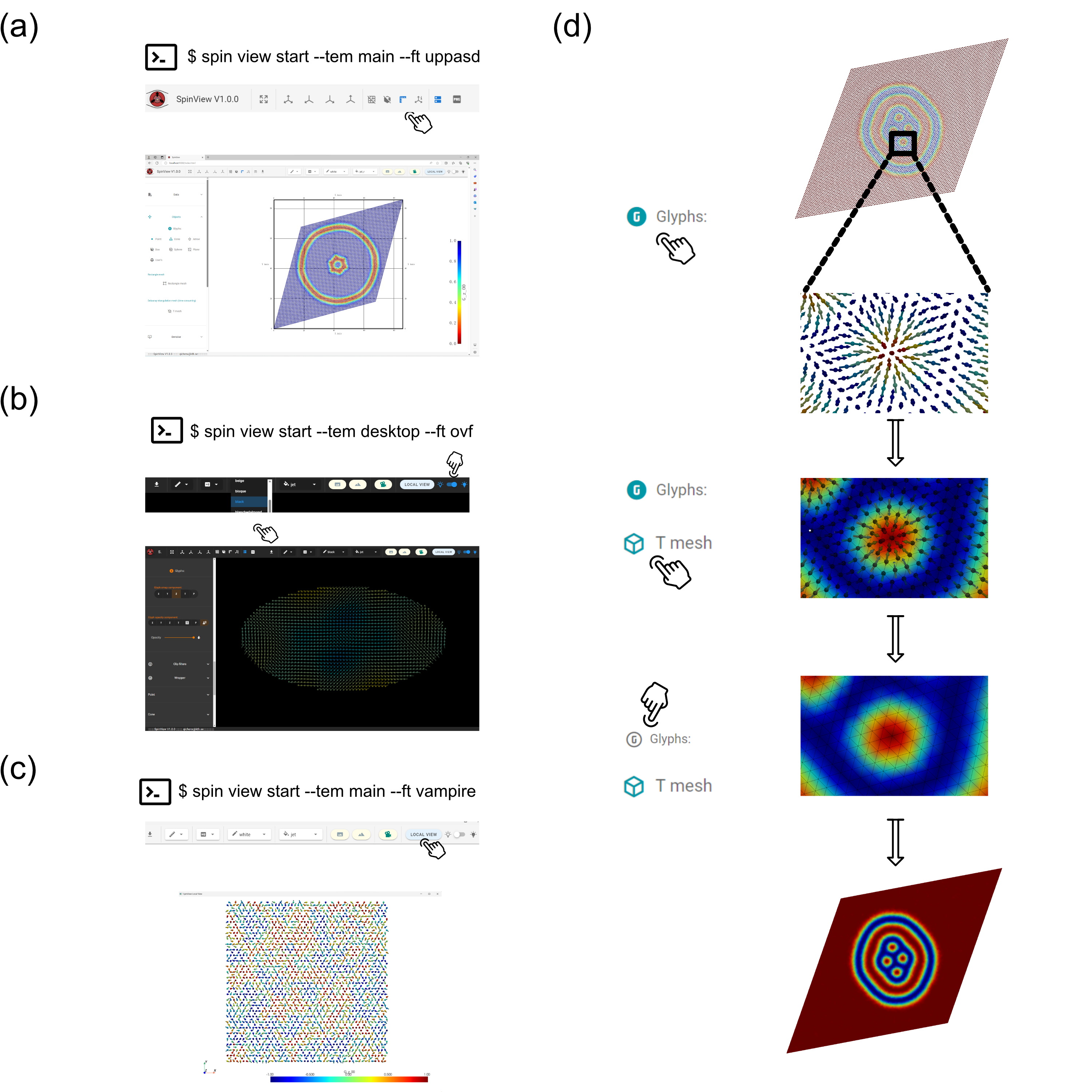}
\caption{Autoparsers for multiple datatypes and the reconstruction of a triangular mesh from a triangular lattice (a) Visualization of UppASD simulation results with the ruler. (b) Visualization of micromagnetic simulation results with special geometry in dark mode (c) Using local view mode to visualize vampire results. (d) Flowchart for switching from using the arrow glyph to the triangular mesh. }\label{secA3_figure}
\end{figure}

\section{}\label{secA4}
The demo of using alloy representation with a built-in SQL database and different kinds of built-in rescale functions is shown in Fig. D3.

\begin{figure}[h]
\centering
\includegraphics[width=1\textwidth]{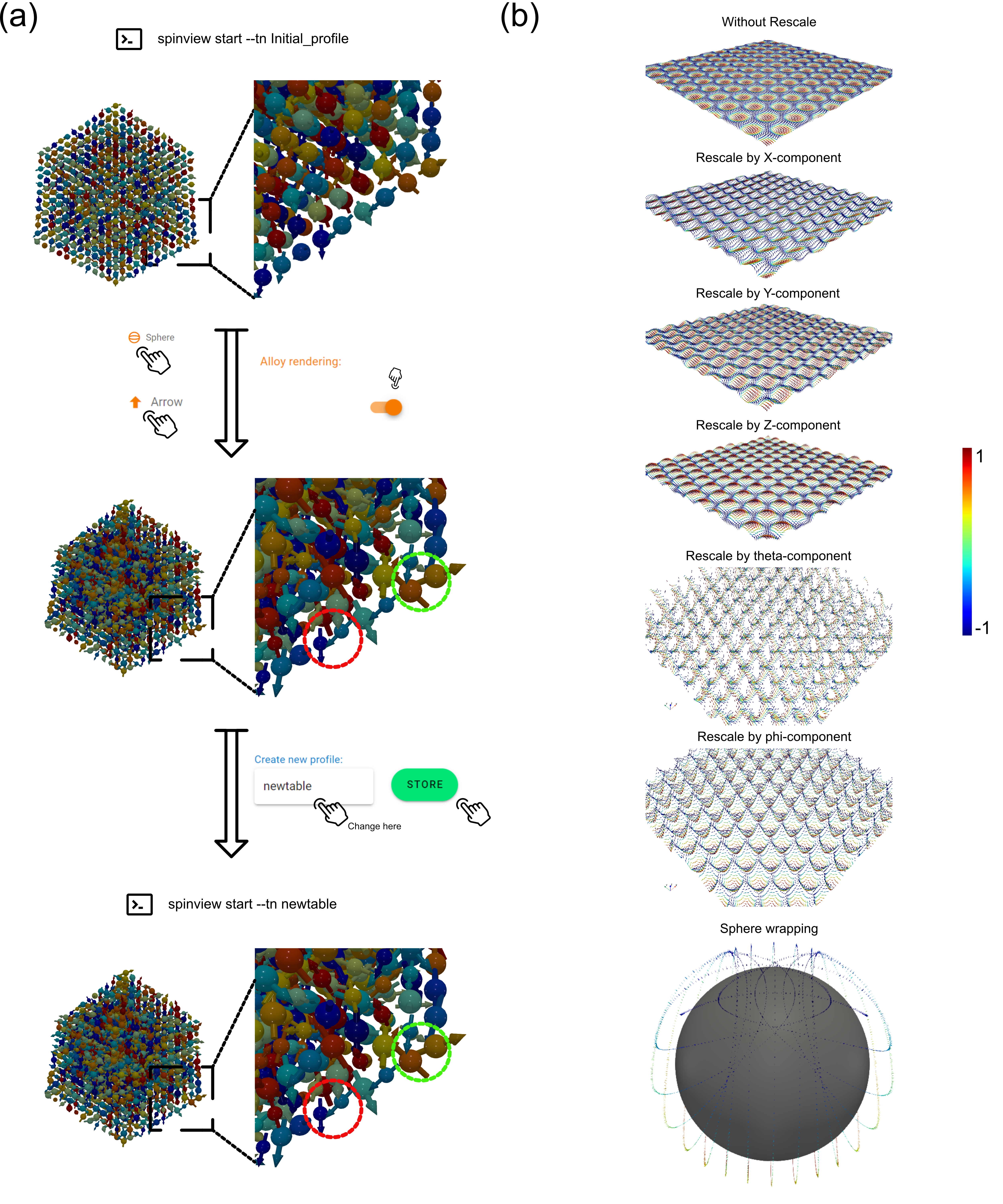}
\caption{Example of using alloy representation with built-in SQL database and different rescale functions (a) Flowchart of applying alloy representation then reusing the setting by storing it in the 'newtable' in the built-in database. The green and red circles highlight the magnitude difference of the magnetic moments of two different atoms in the alloy. (b) Using X, Y, Z, Theta, and Phi components and the sphere warping function to rescale a skyrmion lattice. }\label{secA4_figure}
\end{figure}
\end{appendices}

\clearpage
\bibliography{sn-bibliography}

\end{document}